# Touching the Stars: Improving NASA 3D Printed Data Sets with Blind and Visually Impaired Audiences

Arcand, K.K., Jubett, A., Watzke, M., Price, S., Williamson, K. T. S., Edmonds, P.

(Smithsonian Astrophysical Observatory)

## Abstract

Astronomy has been an inherently visual area of science for millenia, yet a majority of its significant discoveries take place in wavelengths beyond human vision. There are many people, including those with low or no vision, who cannot participate fully in such discoveries if visual media is the primary communication mechanism. Numerous efforts have worked to address equity of accessibility to such knowledge sharing, such as through the creation of three-dimensional (3D) printed data sets. This paper describes progress made through technological and programmatic developments in tactile 3D models using NASA's Chandra X-ray Observatory to improve access to data.

## 1. Introduction

For millennia, astronomy has been a science reliant upon the human eye. Yet in some sense, most wonders in the cosmos would be hidden to all of us without the help of scientific instruments that magnify and vastly expand our collective vision [Rector, Arcand, & Watzke, 2015].

Once the telescope was invented over four hundred years ago, astronomers continued to use visible light, the small range of the electromagnetic spectrum that human eyes can detect, to explore our Universe.

Beginning in the mid-19th century, scientists began to discover that there were other types of light that were invisible without technology. Today we know the full spectrum of light includes radio waves, microwaves, infrared, visible (or optical) light, ultraviolet, X-rays, and gamma rays. Scientists use all of this information to study the cosmos, and recently have been able to add gravitational waves [Abbott et al., 2016] and neutrinos to their tool kits of discovery.

While the origin of astronomical pursuits may have been historically visual in nature, the sense of sight tells only a small part of the cosmic story in modern times. Scientists and science communicators have worked for many years to help tell the tale of the "invisible" Universe through scientific visualizations and translations of data into images that can be processed by the human eye [DePasquale, Arcand, & Edmonds, 2015; Rector, Levay, Frattare, Arcand, & Watzke, 2017].

These have been important developments in helping scientists share their results with the wider scientific community as well as with non-expert populations through visualizations, leveraging a "visual economy" [Bigg & Vanhoutte, 2017, p.118]. However, despite the astronomical wonders perceived in such visual representations, they do not allow for the inclusion of many people.

People with low-vision or no vision who are interested in astronomical discovery may contend with many challenges in processing this information when it is presented in a purely visual form. Such students may be able to use verbal or audio descriptions in order to explore images of data, but it is suggested that these indirect methods are incomplete because blind and visually impaired (BVI) learners must then depend upon secondary knowledge of an image rather than developing an independent interpretation, like their sighted counterparts do as a key component of education [Hasper et al., 2015, p.92].

There is a clear need for improved accessibility to astronomical data in alternate formats that convey the same information as images do, because for participants of all kinds visualizations are an important step in the process of learning [Grice, Christian, Nota, & Greenfield, 2015] and this includes those who are blind and visually impaired [Arcand, Watzke, & De Pree, 2010].

To combat such a deficit, efforts have been underway to transform astronomy outreach materials from a two-dimensional (2D) manifestation to a three-dimensional (3D) representation that can be manipulated digitally as well as physically [Arcand et al., 2017; Madura, 2017]. This helps to establish equity in access to and ability to represent data, since a detailed depiction of the mental representation process illustrates that despite some differences in the BVI population, there are fundamental cognitive structures that produce correct scientific comprehension from perception for all people [Jones & Broadwell, 2008, p.284]. Blind persons are able to gain an understanding of information that is very similar to that of sighted people in regards to phenomena that are normally experienced in a purely visual way, regardless of their different mechanisms of internally representing these concepts [Striem-Amit, Wang, Bi, & Caramazza, A., 2018].

In an interesting paradox, solutions for innovative information sharing methods moving forward can be found by looking back in time through the history of art. Just as astronomy deals with information that can be quite old from a historical perspective on Earth to draw meaningful scientific conclusions, artistic trends from past eras and even prior millenia can inform the design of interactive interfaces today. One time period that featured a strong aesthetic emphasis on multimodal means of sensory communication was the medieval era. The Byzantine culture that flourished in medieval Greece provides an example that is particularly rich in the incorporation of textured surfaces into design [Pentcheva, 2006]. For Byzantine artists the process of sight was comprehended as a tactile activity, and in their creations touch was enabled to work together with other senses such as smell and taste to formulate a holistic model of perception [Pentcheva, 2006].

*"The tangible appeals to and mobilizes all five senses, while the visible addresses itself to just the eye. It is our modern culture's obsession with making things visible, fueled by optical visuality, that makes us project a similar framework onto medieval art."*

Pentcheva, 2006, p.636.

Medieval art made several centuries ago was designed to make use of all of its spectators' senses simultaneously [Pentcheva, 2006]. The modern shift to a primarily visual mode of understanding increased the difficulty of comprehending representations for "non-visual learners" [Grice et al., 2015] and inspired efforts to assist them by adapting technologies to formats that allowed more multisensory interaction. Ever since the eighteenth century Enlightenment, philosophers and other academic figures have sought to understand the experiences of people who are blind or visually impaired [Beck-Winchatz & Riccobono, 2008; Paterson, 2016] in order to comprehend how such learners navigate an often visually oriented world [Skorton, 2018]. Taken together, the thought experiments of sighted scholars such as Descartes, Diderot, and Voltaire worked to help provide an understanding that blind participants use a "sensory substitution" of touch for sight in order to learn independently, a functional shift that has been confirmed by contemporary neuroscience [Paterson, 2016].

The advent of relatively inexpensive 3D printing could allow astronomy demonstrations based on touch to reach wider audiences of the science-interested public around the world. This builds on work by many in recent decades to create representations of astronomical objects in tactile poster form [Grice et al., 2015], as well as in book form [Dawson & Grice, 2005]. For example, tactile astronomy books have been very well received but the cost to print can be high, and therefore the distribution to larger networks can be relatively limited [Beck-Winchatz & Riccobono, 2008; Weferling, 2006].

## 1.1. Legal Framework

The World Health Organization estimates that about 253 million people around the world have vision impairment, with about 36 million individuals who are blind and 217 million people with a moderate to severe level of vision impairment [World Health Organization, 2017]. How does one translate data from a visually-centric field in a way that provides blind and visually impaired persons and populations with direct and equitable access to resources that are often made up of taxpayer-funded information, which all people should be able to understand? One solution lies in moving beyond a determination to focus on the function of sight by transferring information to the domains of the other senses through sensory substitution enabled by innovative strategies.

The Americans with Disabilities Act (ADA), which was passed in 1990, requires that people with disabilities in the United States of America be given equal access and opportunity in regards to a number of public institutions and private businesses. Disability is defined here as "a physical or mental impairment that substantially limits one or more major life activities, a record of such an impairment, or being regarded as having such an impairment", a scope that surely includes blindness and serious visual impairment [ADA National Network, 2019a]. Title I specifies that disabled people should receive the same consideration for employment as all other Americans and reasonable accommodations while on the job that enable them to perform their duties on an equal footing with their colleagues [ADA National Network, 2019b]. Inclusion of BVI people in scientific research and education through fields such as astronomy can help them to understand that professional scientific work is a real option for them, and it can even improve their quality of life [Bonne, Gupta, Krawczyk, & Masters, 2018]. Titles II and III mandate that state and local governments as well as spaces open to the public must be usable and accessible to people with disabilities, with an emphasis on communicating suitably with these populations [ADA National Network, 2019b].

The United States Department of Justice (DOJ) released updated regulations for the Americans with Disabilities Act that applied to cultural and educational institutions in 2010, entitled the *2010 ADA Standards for Accessible Design* [Smithsonian, n.d.]. These guidelines were adopted by the Smithsonian Institution in order to extend its commitment to serving the diverse American audience to the portion of the public with disabilities. The newer rules stipulate that stricter technical standards need to be imposed on new constructions or renovations of government, public, and commercial areas in order to ensure that these spaces are accessible for all people. This commitment of the Smithsonian is also applicable to digital exhibits and other technologies that serve as educational resources to the public, including data. The Institution stresses that although the immediate beneficiaries of these stronger policies aimed at accessibility are the disabled and other socially marginalized groups, the value of more innovative exhibition development strategies will ultimately be a great benefit to all of society due to the creativity developed by presenting information differently in order to engage with groups that had not previously been included in the formulation of design [Smithsonian, n.d.].

Accessibility standards adopted by the United Nations in 2006 with the *Convention on the Rights of Persons with Disabilities (CRPD)* provide even more precise and urgent guidelines on the need to communicate information in formats that are fully accessible to disabled people in an international and global context. Article 21 of this document sets out the conditions for adequate "freedom of expression" for disabled populations, which includes the ability to receive and share information in an equitable manner that is fully comparable to how others engage in these activities [United Nations - Disability, Department of Economic and Social Affairs, 2006].

Article 21 of the *CRPD* is practically applied in the distribution of information and facilitation of communication through methods preferred by and usable for people with a range of different disabilities [United Nations - Disability, Department of Economic and Social Affairs, 2006]. In the case of this project, the particular tools used for enabling access to equitable information sharing are 3D printed models of data sets specialized for BVI audiences, but it is hoped that these efforts will inspire a greater movement towards data accessibility in a variety of mediums.

1.2. Audience and Approach

When considering the particular needs of BVI audiences, especially children, it is important to consider that such learners greatly depend on individualized support for the development of specific skill sets [Bonne et al., 2018; Raisamo et al., 2006]. It is imperative to help participants in testing feel engaged, working with them as partners in the development of inclusive resources through the design of interfaces that build upon their different abilities [Skorton, 2018] in order to draw from a diversity of approaches to data representation that enriches the toolkit of science [Schmelz, 2015]. This multimodal paradigm broadens the framework of participation for underserved groups.

Defining equity and access in relation to 3D printing resources for BVI audiences can help to clarify the importance of creative and personalized communications with this population. Regarding access to information, *equity* refers to intended recipients encountering data that leads to an experience of learning perceived as valuable and suitable to their unique contexts, rather than simply their ability to obtain a great volume of data that may come across as indecipherable or irrelevant [Lievrouw & Farb, 2003]. James E. Porter's 1998 book *Rhetorical Ethics and Internetworked Writing* establishes metrics for *access* on the part of underrepresented groups that should include elements of infrastructure such as physical spaces and technologies acting as entry points, literacy through the ability to interpret information, and integration into the broader community [as cited in Dawson, 2014]. It is important to consider the motivations and needs of the targeted audiences in all aspects of interface design when applying these criteria to informal science activities, especially when seeking to develop inclusive practices aimed at underserved populations [Dawson, 2014].

MakerSpaces - public facilities with access to tools like 3D printers - and related settings can help emphasize creative participation in the learning process that increases ownership of the material through critical thinking [Giannakos, Divitini, Iversen, & Koulouris, 2015], and the incorporation of accessible technology in group settings has been shown to increase the engagement and confidence of people with disabilities [Nosek, Robinson-Whelen, Hughes, & Nosek, 2016]. Finding innovative ways to organize, share, and analyze data with various techniques that make use of multiple senses [Díaz-Merced, 2014] can reveal details that could potentially be missed by focusing on a single sensory modality. This method would enhance the quality of knowledge obtained from the interpretation of results for the scientific community as a whole, BVI and sighted members alike.

The need to integrate information from a variety of different sensory domains illustrates the importance and relevance of *universal design* ADA National Network, 2019a; Childers, Watson, Jones, Williamson, & Hoette, 2015] which creates formats that are both usable and accessible for any participant, regardless of ability. The roughly 100,000 young people who are blind in the United States alone [Beck-Winchatz & Riccobono, 2008] could greatly benefit from educational programming designed to prioritize them on equal footing with their sighted peers [Bonne et al., 2018].

There have been a variety of unique sensory programs in astronomy developed in recent years, from some designed around sound (e.g., "Walk on the Sun" from the Design Rhythmics Sonification Research Lab [Quinn, 2012], and research by astronomer and computer scientist Dr. Wanda Díaz-Merced [2013]), to others built with a focus on taste [Trotta, 2018], smell [Wenz, 2018], and vibration [DeLeo-Winkler et al., 2019]. This paper will concentrate on the sensation of touch, and the creation of 3D prints to help serve a need for specific kinds of access incorporating that modality.

1.3. Technique

Research from a number of studies with people who are blind or visually impaired suggests that the visual cortex perhaps performs more multi-sensory or multimodal and spatial functions than its name implies [Jones & Broadwell, 2008]. Past strategies of communicating astronomical ideas through touch have included employing marbles, a knotted cord, and grains of sand to model proportions in the Universe [Weferling, 2006] and using cardboard and wire to make the phases of the Moon tangible [Alonso, Pantoja, Isidro, & Bartus, 2008]. More recently, scientists have found 3D prints to be an effective format for communicating the shapes of astronomical objects such as Eta Carinae [Madura, 2017] and the Moon and planets in the Solar System [Jones & Gelderman, 2018] while turning their component parts into tactile models, clarifying details of their structures.

3D printed versions of astronomical data with tactile features have been shown to help participants who are blind, both those who are blind from birth and those who have lost sight at some time after birth. From "stimulating, building, and reinforcing a person's mental model" of the objects to self-reported comprehension and learning gains, positive outcomes were reported, including the ability for some participants to visualize the data being modeled [Christian, Nota, Greenfield, Grice, & Shaheen, 2015, p.43]. Building spatial reasoning skills has been shown to be very important for success in science, technology, engineering, and math fields but is often less developed in underrepresented groups [Jones & Broadwell, 2008].

Non-expert populations can benefit from tactile or 3D-printed models of science data, including students [Hasper et al., 2015] and populations with visual impairments [Christian et al., 2015; Grice et al., 2015]. Additionally, 3D printing is able to address distribution concerns because it has become more commonplace in recent years, with 3D printers situated in many local U.S. libraries, schools, MakerSpaces and more [Childers et al., 2015; Grice et al., 2015; Madura, 2017].

1.4. Overview of the 3D astronomical data sets

Obtaining three-dimensional information about astronomical images can be quite challenging. Astronomers cannot normally fly around objects with their instruments. A small number of Solar System sources that spacecraft have encircled, such as Saturn which was viewed in its totality by the now-retired Cassini spacecraft of the National Aeronautics and Space Administration (NASA) [NASA Science, 2011], serve as the exception to this principle. In order to explore such astronomical objects in three dimensions, special techniques and even serendipitous circumstances must be utilized (see for example descriptions in Arcand et al. [2017], DeLaney et al. [2010], and Orlando et al. [2017]).

The examples from this evaluation feature 3D data from primarily NASA telescopes that were gathered on three separate stars or star systems, two of which are the debris fields of exploded stars. While other astronomical objects do lend themselves to being studied in 3D, stars and supernova remnants have become a particular focus for the Chandra X-ray Center (CXC).

There are several reasons for this, and the most important among them are that supernova remnants are often very bright in X-rays, some are relatively close, and they are changing on human time scales. They also contain a large amount of structural detail and variety. These factors and others have allowed the Chandra team to gather actual data that can be translated into a 3D form, thus avoiding the need to estimate or imagine what the object looks like beyond a 2D image seen on the sky [Arcand et al., 2017].

The manipulation of scientific data, whether into 2D or 3D form, can lead to a loss of certain information. Not all information can always be included in a data output, regardless of the number of dimensions involved. Such information loss is not necessarily a negative. We share an emphasis on communicating current astronomy data in a realistic manner with the recent Tactile Universe project in the United Kingdom [Bonne et al., 2018]. Data selection is part of the pipeline of turning binary code received from telescopes into forms able to be processed by the general public [Arcand et al., 2013], and it can help highlight or focus on useful features [Madura, 2017]. An example of this in 3D printing is that small pieces of the 3D data models that are not attached to any material near them would have to be deleted (removing of data) or forcibly attached (adding of data not there) in order for the model to be printed as a single object [Arcand et al., 2017].

The next section of this paper describes the latest efforts at NASA's Chandra X-ray Center to fuse the advances in 3D printing and 3D datasets to literally put astronomical objects into the hands of new audiences, especially BVI communities, for exploration and education.

## 2. Methodology and Evaluation

Overall, for the Chandra-based 3D print models and Braille program, the authors have worked with numerous groups for informal and pilot testing of the models since 2015. These include state-wide-BVI initiatives such as the Rhode Island College BVI program, federal programs such as the NASA 508 committee of US federal employees, international programs such as the International Astronomical Union's Inclusive Astronomy committee, and programs for formative testing and accessing dissemination networks. Based upon the formative testing, the authors chose three discrete astronomical objects (Cassiopeia A, Supernova 1987A, and V745 Sco).

Formal testing of the three 3D printed models was completed with two groups of students, totaling 19 participants ages 14-21, at the Youth Slam for the National Federation of the Blind (NFB), from July 26-27, 2017 with Institutional Review Board (IRB) approval accepted through the National Federation of the Blind. Three presenters from the authors' home institution were present at the session, with one presenter describing the majority of science and object types and the two additional presenters assisting in the handling, discussion, and gathering of feedback.

The following section reports on the interactions and responses to the 3D objects once in the hands of the students at the National Federation of the Blind conference (including the information that was explained to them, specific reactions to the models, and example questions that arose). Discussions were recorded in audio format with user and organizational permissions, with additional notes taken of the interactions recorded by the presenters.

### 2.1. Cassiopeia A:

The two Cassiopeia A (Cas A) models shown to the participants were 3D-printed using two types of filament. 3D printer filament for Fused Deposition Modeling (FDM) comes on spools as thin ropes of thermoplastic material, usually around 2 mm in diameter. The composition of the plastic varies and is selected by the user based on properties such as strength, flexibility, melting temperature and renewability versus cost. Polylactic Acid (PLA) filament is a biodegradable plastic made from cornstarch and other natural materials. The filament is fed into the printer through a print head. Print heads in the 3D printer heat up and melt the filament, which is extruded in a pattern specified by digitally modeling and slicing into layers. The model builds up, layer by layer, on the print bed. PolyVinyl Alcohol (PVA) filament, the same material used in laundry detergent pods, is a dissolvable filament used by some printers with dual extruders to build support structures as the model prints, supporting overhanging parts of the model as the molten PLA filaments harden in place. The PVA is extruded from a second print head in the same way that the PLA is extruded, layer by layer onto the print bed. The entire model is dunked in water after printing, and the PVA supports dissolve away, leaving only the PLA model.

The full model of Cas A was about 10 cm in diameter and took approximately 32 hours to print. A second model, a cross-sectional version, consisted of two pieces that fit together using magnets. Each half, about 13 cm across, took about 24 hours to print.
The presenters explained to the participants in two group settings that light from the supernova explosion reached Earth some 300 years ago, and that the object is physically 10,000 light years away in our own Milky Way galaxy. The group discussed what a light year is, as a measure of distance and what that means.

Next, the two copies of the 3D printed object were distributed for everyone to touch and explore. The first model was the typical 3D Cas A (figure 1a), the second model of Cas A consisted of the two halves split apart revealing the shell-like shape (figure 1b). Presenters explained that the spiny features are jets and that 3D data tell us that Cas A exploded in two waves, the first more or less spherical and the second with the jets protruding outward. Such information is interesting to scientists because the 2D data of this object make Cas A appear very circular. This novel representation therefore influences how scientists approach the creation of new numerical models of supernovae and supernova remnants.

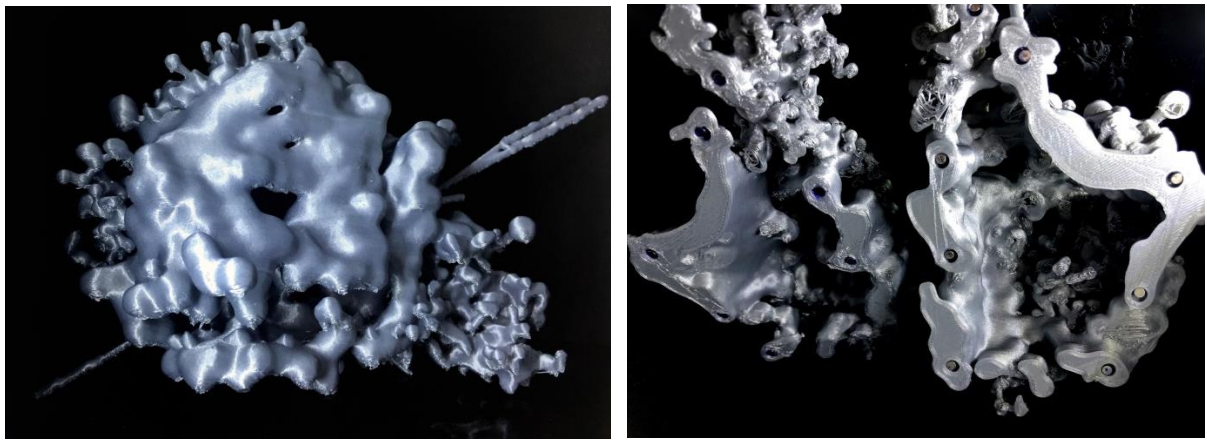

*Figure 1 a & b. 3D model of Cas A via Ultimaker 3 in single (left, a) and dual part (right, b) versions. Visit [http://chandra.si.edu/3dprint/](http://chandra.si.edu/3dprint/) for model files. Credit: NASA/CXC/K.Arcand*

The cross-sectional model, which had been in development specifically for blind and visually impaired audiences, allowed participants to explore both the internal and external data of the object, revealing its shell-like shape and interior neutron star. Touching the hollow interior of this model can help participants to better comprehend the appearance of Cas A and distinguish the tactile surface from the distant object it represents, and therefore it helps to achieve a stronger sense of the 3D printed object's scale in relation to the supernova remnant that served as its model [Pentcheva, 2006].

Using a qualitative approach, students were verbally asked a series of questions about each model as they were handling them. Questions asked of participants included:

*-What did you think of the 3D printed materials you used?*
*-Are there aspects which could be improved?*
*-What do you feel you've learned from these models?*
*-What would you change with how they are presented?*
*-Has working with the 3D printed models changed how you relate to astronomy or how you view yourself in relation to science?*

Their verbal responses to the formal questions as well as impromptu remarks were recorded.

*Table 1. Categorization of general comments into blocked themes.*

| What did you think of the 3D printed materials you used this week? | Are there aspects which could be improved? | What do you feel you've learned from these models? | What, if anything, would you change about how they are presented? |
|---|---|---|---|
| "I thought they were really cool!" | "More print materials would be nice." | "I learned what stars looked like when they exploded." | "More models" |
| "Neat and helpful." | "Size better helps feel, make larger." | "How stars die and if they want to, blow up and "vomit hot gas." | "Larger." |
| "I thought the class was ok. but I would of [*sic*] liked more a focus on the 3d printing process, rather than on space. Which was a subject I already knew a lot about." | | "That a stars shape is more bumpy than I thought." | "I would make more models to show more stars. For instance, I would make models of consolations [*sic*] to show those." |

Specific comments captured on Cas A from participants included:

*-"I thought it would be more like something splitting, not puffing out."*

*-"It's more prickly than I expected." (This resulted in a discussion about how if we're not careful about giving all the information, a model of a supernova remnant made out of solid plastic can be pretty misleading since the data are representing mostly gas and dust and since the shape of the data is always changing).*

*-"I imagined a ball."*

*-"It's kind of spherical, but it has a huge tail."*

The participants explored the shape and size of the objects, comparing them to objects or shapes they were already familiar with or expecting to encounter. Participants were, at points, worried about breaking the models, though none were broken in this event. The two different Cas A models were different sizes, and that seemed to make a difference, especially in the feeling of the model being "prickly".

## 2.2. Supernova 1987A

The Supernova 1987A (SN 1987A) model looks and feels quite different from the Cas A model the participants had already worked with (figure 2). SN 1987A shows how the material from the supernova explosion expanded and pushed through an existing ring of material from a previous eruption event by the progenitor star to create a crown-like shape with fingers of high-energy material perpendicular from the ring. One SN 1987A 3D print was produced using PLA/PVA filament on the Ultimaker 3, taking approximately 32 hours to print at about 8 cm in diameter. A second copy of the same SN 1987A model was printed on the Ultimaker 2 (a less expensive printer compared to the Ultimaker 3), a process which requires hand-removal of the support structures and results in a bumpier, more jagged and less smooth 3D print. Participants noted the jaggedness in some of their remarks.

A short discussion was held about the SN 1987A object and the basic premise of its science with the participants. Scientists have been collecting data on SN 1987A since the explosion was detected in 1987, and Chandra has been keeping track of its changes every year since the observatory launched in 1999. We now have 3D data on this supernova remnant as it changes over time by using the X-ray observations to constrain models of the explosion. We selected a single epoch (2017) for the 3D printed model out of the almost two decades of data. At its current state, the ejecta from the explosion has nearly completely passed through the original gas ring and is starting to disperse. The matter will eventually no longer be recognizable as a cohesive shape. Depiction of this changing state requires cooperation between the senses [Pentcheva, 2006].

Specific comments captured on the SN 1987A model from respondents included:

*-"It's a ring instead of a popcorn puff."*

*-"Like a crown or a basketball hoop."*

*-"Space is weird."*

*-"What kinds of choices do you have to make when representing data?"*

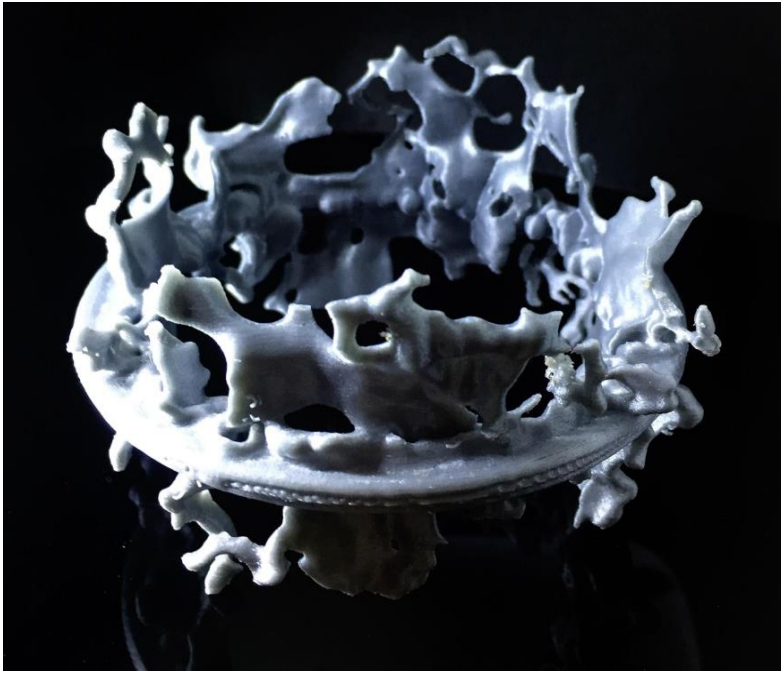

*Figure 2. 3D model of SN1987a via Ultimaker 3. Visit http://chandra.si.edu/3dprint/ for model files. Credit: NASA/CXC/K.Arcand*

Participant comments sparked a larger discussion about how Chandra detects photons. We described that data starts first as numbers showing 2D data, then are translated as x and y positions on a grid plus the energy level of each photon. Next the Doppler effect (when possible with the data) can be used to explore the red- and blue-shift (the motion away or toward the viewer) of each data point to determine the 3D shape. Scientists can take the points, now mapped to 3 axes (x, y, z), plot them in a computer program and wrap a virtual mesh around the data.

Decisions need to be made about how thick to make the mesh and how much data to encase. This brought up the changing data that was discussed regarding 18 years of SN 1987A observations and how the early data look a bit like scattered blobs floating in space. Earlier data from 1999 or 2000 would have been quite plain - presenting as a simple ring - in comparison with the present day epoch (then 2017), so our research team decided not to prioritize earlier SN 1987A data for 3D printing at that time.

The presenters further talked with the respondents about how this data visualization or selection poses an interesting challenge. Figuring out how to best tell the story while being as true to the data as possible is a crucial issue in any kind of scientific visualization. It was noted that it is important to remember that the astronomical objects are dynamic, as the gas and dust are dissipating and moving at thousands of miles per hour through intergalactic space. The presenters also discussed the elements being created first within the stellar core, then in the actual explosion, and then moving out and seeding the Universe with elements essential to life.

2.3. V745 Sco

The V745 Sco model was printed using two colors of PLA/PVA filament to differentiate the separate parts for sighted audiences. Pieces printed separately will tell this same story to those feeling the models for details. The ejecta took about 20 hours to print, the shockwave about 24 hours, and the ejecta embedded within the shockwave took close to 40 hours. The complete model is about 10 cm wide and represents a cross-section of the nova, a binary star system that undergoes irregular outbursts.

This model tells its story best when presented in 3 parts, the ejecta, the shockwave, and then both parts together (figures 3 a, b, c). As a result these models were passed around as a set, accompanied by the explanation that the bumpy one is the matter ejected during a nova and the smooth part is the shockwave created during the same event. The students were asked if they could tell that the 2 distinct parts could fit together, one inside the other, and whether they could tell more of the story with these parts separated.

Specific comments captured regarding the V745 Sco model included:

> -"It looks like it could fit, but I know it doesn't."

The researchers spoke about how the pieces couldn't be retrofitted together, based on the fact that the models are solid plastic. Instead, there was a discussion about trying to imagine a softer material being pushed into the space of the other, or gas and dust filling the space and deforming the shape of the shockwave.

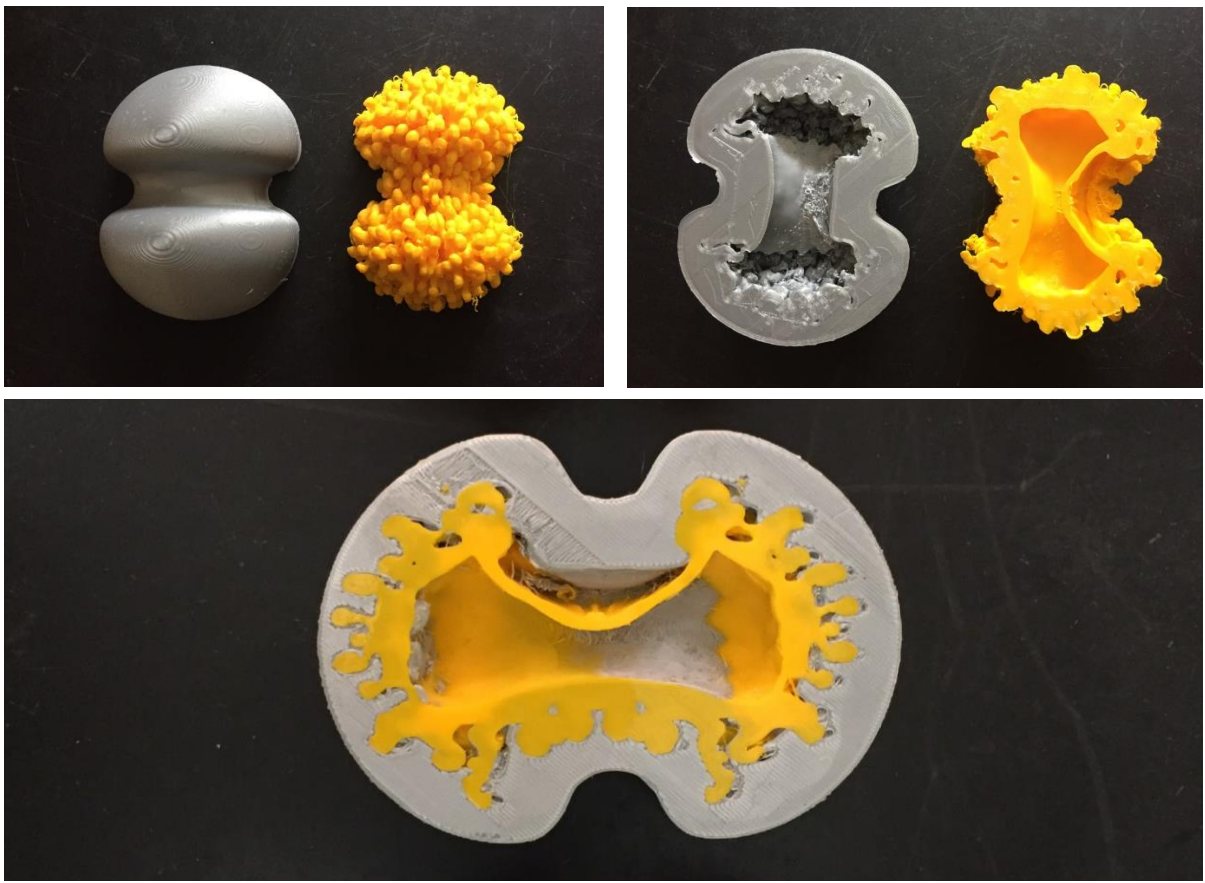

*Figure 3 a b c. 3D models of V745 Sco exterior (a) and interior (b) (shockwave on left, ejecta on right in each, shown in cross-section) via Ultimaker 3. The version showing the joined model (the difference only appears in the interior) is shown in c. Visit http://chandra.si.edu/3dprint/ for model files. Credit: NASA/CXC/A.Jubett*

One participant suggested using a polymer gel and another spoke about silly putty to make a softer model that could be pushed into and pulled out of the space in the shockwave as alternative methods.

## 2.4. Overall Feedback and Response

Overall, the feedback from participants on the models was very positive (refer to table 1 for categorized comments). There were useful discussions at the event on size, texture, quality and quantity of the 3D prints. In previous BVI projects the first author was involved in [Arcand et al., 2010], experiments were performed with the scale of Braille and tactile image panels for an exhibit. Participants who were BVI noted in these projects that the panels were too large (60 cm across) and would be better at a legal size pad. The pilot project reported here was conservative with the scales of the 3D prints, though numerous participants expressed the desire for larger 3D prints to be able to access the information in greater detail.

It is indeed possible to print some of the models in a larger format — for example, producing Cas A at about 30 cm across to good effect on an Ultimaker 3 takes about 52 hours to print. However, it should be noted that those larger sizes are harder to print, can malfunction easier during the print process and are much more expensive and time consuming. When feasible, therefore, it is recommended to print larger models for BVI audiences, but within the constraints of the product budget and timeline.

As mentioned above, resolution and print quality are important aspects to consider. Respondents commented specifically on model textures, patches of roughness and areas of prickliness that did not seem to flow with the rest of the object. This was noted mostly in the prints where the removal of support structures had been done manually. When touch is the main sense being used to access these 3D models, particular care should be taken to provide the best quality of data possible.

It is essential to pay special attention to texture differences that may emerge during the printing process in order to help prevent data from being misread by tactile learners. For example, blind individuals might misunderstand artifacts or bumps in the models from poor support structure removal or lower resolution printing. These variances can be misleading for any audience, but they are particularly disruptive for the BVI demographic. Using a dissolvable substrate printing process can help alleviate the worst of the artifacts from manual removal of support structures, but again, cost may be a prohibitive factor for some venues.

Other general points included that respondents would have appreciated more models to explore. It is possible to augment a smaller collection of astronomical models such as the ones presented in this specific instance with additional models of simpler objects to introduce the subject matter at hand.

For example, in other pilot events the authors have included data-based 3D printed models of the Moon and Mars as a starting point for discussion of astronomy generally, then moved on to models of stellar formation (Eagle Nebula/"Pillars of Creation") and mature stars (Eta Carinae) to talk about other steps in the stellar evolution cycle. Those models were accessed from other NASA observatory data sets and are collected on the NASA 3D archive (https://nasa3d.arc.nasa.gov/models). As more groups develop 3D models it should be easier to access supplementary materials such as these.

## 3. Results & Discussion

Evidence from self-selected qualitative studies at the NFB events, along with anecdotal information in informal testing and individual recommendations from users all suggest that the 3D printed models test very well for their intended purposes. Minor adjustments were made to the 3D prints based directly on user feedback. Examples include adjusting size, reinforcing various parts to be forced into a sturdier size/connection (e.g., a jet that was too fragile), and paying attention to printer resolution and "cleanliness" of the resulting model. One of the primary conclusions was that cutting more complex and "whole" 3D models in half worked very well (for both non-sighted and sighted audiences), to provide internal and external access to the information and help better inform the creation of the mental map of the object.

One potential concern with printing solid static models to represent astronomical objects that are constantly changing and dispersing, and that often primarily consist of gas and dust, is how the use of these plastic, hardened materials could be misleading or lead to misconceptions. One of the reasons the cross section helped convey the science better is that it showed the material as a shell rather than as a perceived solid object. Additionally, it also exposed the mostly cleared out area around the neutron star and helped provide a more complete and accurate story of the star's death. More research is necessary to discover what misconceptions may be arising by using physical models in this manner.

One issue touched upon with the participants, and which has been discussed at other events since, has been the issue of scale. The stars being held by the participants are greatly scaled down from their actual size. Reference points to scale are typically provided during any discussion of the objects and in any supplementary audio or Braille materials.

For example, the surface area of Cas A is about $2 \times 10^{20}$ times the surface area of the Earth. Therefore, a 3D printed model of Cas A at 10 cm has been scaled down many orders of magnitude. Further research should be done to explore whether the jump from 3D printed size to actual size is not clear, or if other misconceptions are being created in this exploration through 3D print, and if that is the case, what else can be done to alleviate such issues.

The authors have also continued to experiment using magnetized parts to allay some of the fragility of certain aspects of the models, to good effect. For example, the jets protruding in opposite directions from the Cas A remnant have had a high incidence of breaking off during use and shipments. Given the time that it takes to print a single Cas A model (see examples provided in previous sections), tactics were devised to avoid printing a new model each time the jets broke. Losing the jets leads to a removal of an important part of the story of how Cas A exploded in two waves, so it was an important issue to address. A solution to the delicacy of those specific parts has been to print the jets separately and glue small but strong magnets to the edges for detachment during storage and shipping, and then reattach them for use (figure 4).

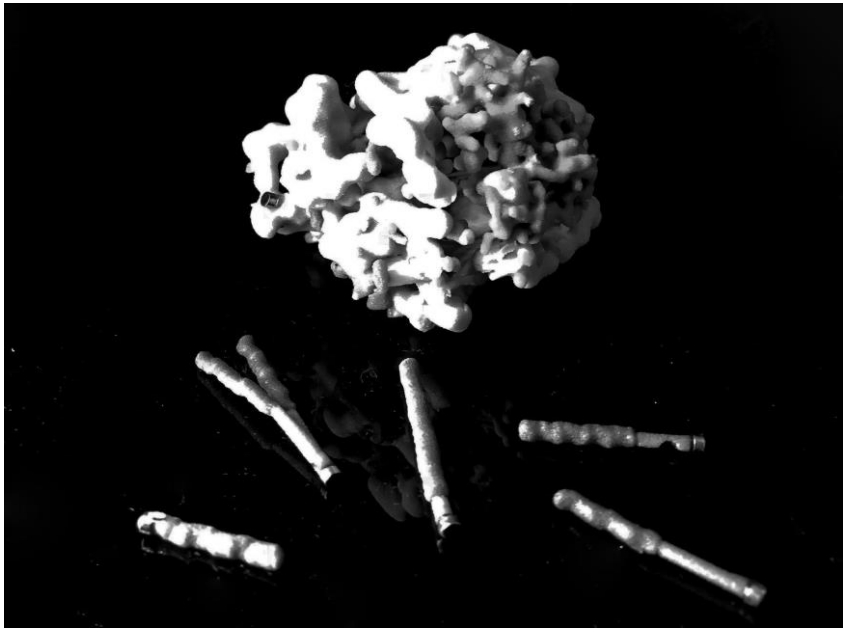

*Figure 4. 3D model of Cas A with magnetized jet, disassembled to show the magnets. (Please refer to figure 1 to see how the jets fit). Credit: NASA/CXC/K.Arcand*

With this technique, when a jet breaks, it is possible to reprint only that single small part (which typically takes less than an hour). It is important to note that the jets themselves break at a much slower rate than when these jets are adhered non-magnetically to the supernova remnant as a whole. (Caution should be applied to the use of small, magnetized parts if working with very young children. In our cases, the respondents were teenaged or older).

## 4. Next Steps

The authors' next step in the process of helping participants to touch the stars is the creation of a complete Braille/tactile and 3D printed kit for dissemination to BVI events, libraries, and schools. This kit (figure 5) consists of five 3D prints, including Cas A, SN1987A, V745 Sco, Eta Carinae, and the Pillars of Creation. Provided in the box are a selection of tactile and Braille panels on general, introductory astronomy topics (explosions, shadows, spin of objects such as black holes or galaxies, seeding of the interstellar medium with chemical elements from supernova remnants, spirals, and wind), as well as a Universal Serial Bus (USB) drive preloaded with audio files describing each of the five 3D-printed objects included, and audio files of the text from the Braille panels. The tactile and Braille panels (http://hte.si.edu/tactile.html) were included in the kit after previous evaluations (unpublished) of astronomy content for audiences with visual impairments as well as direct user requests asked for introductory science content to be provided in order to help lay the foundation for the astronomical topics being covered. The cover of the box is printed in Braille, and a Braille map of the box's contents is also included.

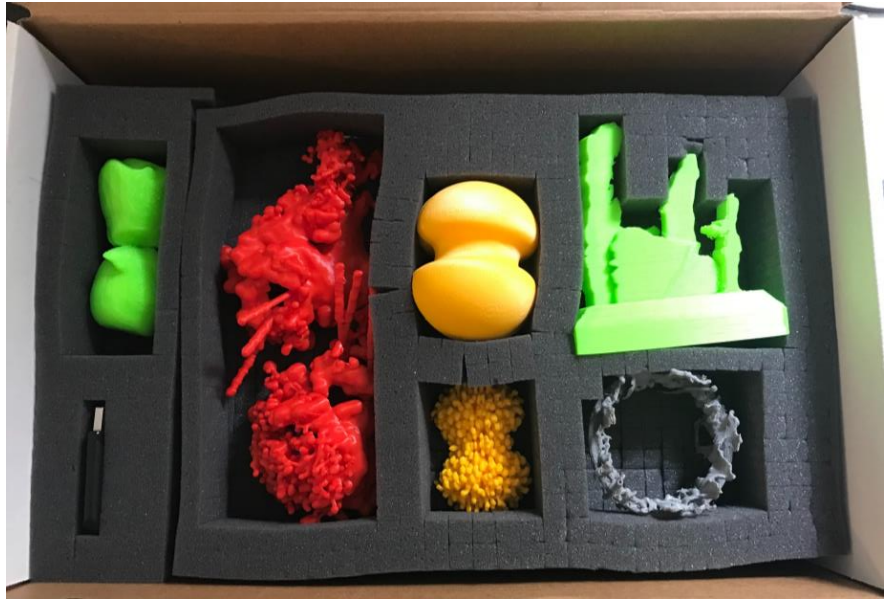

*Figure 5. Pilot of 3D print and Braille box showing the first level of 3D prints. The kit includes a Braille cover, Braille (and audio) information sheets and Braille/tactile posters (not shown). Visit http://chandra.si.edu/tactile/ for details. Credit: NASA/CXC/K.DiVona*

Further testing and piloting around the BVI kit are currently being conducted with the International Astronomical Union, the NASA 508 Accessibility group, and at additional events with BVI audiences at NASA centers including the Johnson Space Center and Marshall Space Flight Center.

Simultaneous work is being completed on further refinement of the models with slight simplifications so that they can be printed through commercial 3D printing companies to help increase production quantities, while also providing slightly simplified models for others engaged with 3D printing including librarians, educators, and people in makerspaces and informal learning venues. The intention is to expand dissemination plans and networks, as well as move the creation of new data-driven 3D printed models forward. The creation of widely available channels for distributing comprehensive BVI kits is an important component of overall accessibility for this population [Bonne et al., 2018].

Finally, the team has acquired a powder printer from a commercial printing company to produce models in full color, and has developed a model of Cas A with color coded information intact to denote the chemical and/or electromagnetic spectrum-based makeup of the remnant in various areas. The powder printer layers gypsum powder (a cornstarch-like material) on the print bed and injects ink from an inkjet print head and glue according to the digital slice and its color information to hold the model together. The untouched powder is vacuumed back into the machine after the print is finished, and the resulting object is processed further to strengthen the full-color model.

One future goal of the project is to apply unique textures assigned to each color to provide the additional color information in a tactile format for participants who are visually impaired, which is crucial for such audiences to make distinctions [Jones & Broadwell, 2008; Madura, 2017] and is a key concept in galaxy and star evolution [Bonne et al., 2018]. Recent research has shown that blind and visually impaired individuals process color in a similar manner to the way that sighted people comprehend language because they experience it as an abstract phenomenon lacking external reference points [Striem-Amit et al., 2018], and so it is important to find a way to translate this information with tactile features.

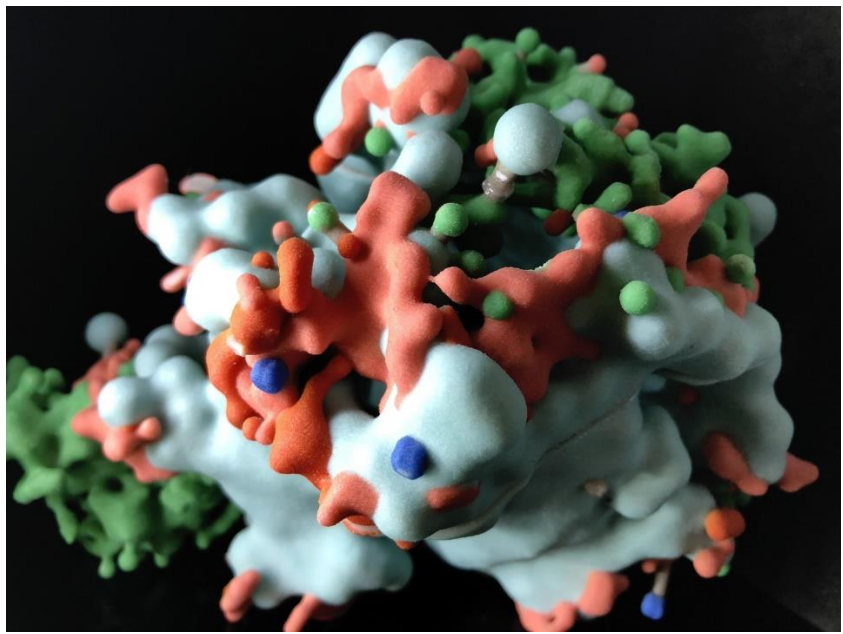

*Figure 6. 3D model of Cas A via the powder printer process allows the preservation of the multi-colored aspects where each color represents a chemical element and/or energy. Credit: NASA/CXC/K.Arcand*

## 5. Conclusion

Enabling access to knowledge within the context of science, particularly government-funded astronomy and astrophysics in this specific case, is crucial in order to provide BVI and other differently abled participants an equitable and equal opportunity in the data sharing ecosystem. Such products may take the form of tactile 3D printed models, virtual reality (VR) experiences with audio capabilities, and data sonifications, as well as other products that engage different senses.

For the purposes of this paper, the focus was placed on 3D printed technology of NASA data-driven 3D models as a cost-effective, distributable product to provide the collected data in an accessible form. The key to the 3D printing program presented in this paper, however, has been to work closely with participants who are blind or visually impaired to directly learn what techniques and approaches work best.

Astronomy has moved into a new era of "multi-messenger" data, where information from all types of light are now joined with gravitational waves and neutrino detections. Astronomy communication may also be poised to enter into a new phase. We have shown that the days of solely sharing the wonders of cosmos in two (visual) dimensions are behind us. Instead, we look forward to exploring space with audiences in as many formats and sensory experiences as possible. Projects involving 3D datasets and 3D printing may well represent one such future of communicating astronomy with the public. It is a future that we are already able to reach out and touch.

## Acknowledgements

Special thanks to Natalie Shaheen and the NFB for inviting the authors to participate in the NFB YouthSlam event. Gratitude also to the Smithsonian Digitization and 3D Lab in Washington, D.C., which inspired the first author to pursue 3D printing the first ever data-driven supernova remnant and set us on the course of using 3D prints to work with new audiences. The Cassiopeia A digital 3D map was originally developed with Dr. Tracey DeLaney of West Virginia Wesleyan College, and the SN 1987A and V745 Sco models were developed with Dr. Sal Orlando of the National Institute of Astrophysics in Rome, Italy, in partnership with the Chandra X-ray Center at the Smithsonian Astrophysical Observatory, in Cambridge, MA, with funding by NASA under contract NAS8-03060. Many thanks to those researchers. Portions of this paper were presented at the 2018 American Geophysical Union (AGU) Fall Meeting, 10-14 December in Washington, D.C.

## Disclaimer

The Chandra X-ray Center (CXC) does not endorse any specific commercial product.